\documentclass[pdflatex,sn-standardnature]{sn-jnl}

\usepackage{amsmath}
\usepackage{lineno}
\usepackage{natbib}
\usepackage{mathtools}
\usepackage[section]{placeins}
\usepackage{textcomp}

\jyear{2024}%

\theoremstyle{thmstyleone}%
\theoremstyle{thmstyletwo}%

\theoremstyle{thmstylethree}%

\begin{document}

\title[Optical solitons in multimode fibers: recent advances]
{Optical solitons in multimode fibers: recent advances}

\author*[1]{\fnm{Mario} \sur{Zitelli}}\email{mario.zitelli@uniroma1.it}

\affil*[1]{\orgdiv{Department of Information Engineering, Electronics and Telecommunications}, \orgname{Università degli Studi di Roma Sapienza}, \orgaddress{\street{Via Eudossiana 18}, \city{Rome}, \postcode{00184}, \state{RM}, \country{Italy}}}




\abstract{Optical solitons in multimode fibers have been predicted 40 years ago and extensively investigated theoretically. Transmission experiments in nonlinear multimode fibers have gained renewed interest, motivated by their potential to extend the capacity of long-distance transmission systems; only in the last few years, new experiments have revealed unexpected properties of optical solitons propagating in graded-index and step-index multimode fibers, partially re-writing the existing theory.
Here we provide an overview of the recent experimental, numerical and theoretical studies which revealed those new properties. It will be shown that multimode fiber solitons form with specific pulse width and energy dependent on the wavelength, and that they naturally evolve toward fundamental-mode Raman solitons. New soliton fission mechanisms, governed by the modal dispersion, will be explained. Possible applications in space-division multiplexed systems will be discussed. A recent thermodynamic approach to soliton condensation will be described.}\footnote{The license does not apply to third-party figures in this article, which remains subject to the terms of the respective rights holder.}


\maketitle

\section{Introduction}

Solitons are stable solutions of the nonlinear Schr{\"o}dinger equation (NLSE), governing the wave propagation in single-mode as well in multimode fibers (MM).

Historically, MM fibers have been used in the 1970s and 1980s, before the development of single-mode fibers \cite{6772470}, in telecom systems of the first generation, which made use of LED sources and GaAs semiconductor lasers at 850 nm wavelength, and of the second generation, which used InGaAsP lasers operating at 1300 nm. The cross-talk among modes, together with the modal delay responsible for temporal delays among modes, severely limited the bandwidth of such systems to below 100 Mb/s. Hence, MM fibers were soon replaced by the single-mode fibers which extended the channel capacity to 2 Gbit/s already in the early 1980s. During the first decade of 2000s, optical transport networks based on single-mode fibers adopted advanced technologies, such as wavelength-division multiplexing (WDM), coherent detection, advanced amplitude and phase modulation formats, Raman amplification and digital signal processing (DPS) for chromatic and polarization dispersion compensation \cite{doi:https://doi.org/10.1002/0471221147.ch5}. Hence, the Shannon capacity limit for a single mode fiber channel was reached \cite{5420239}. The availability of new multiple-input multiple-output (MIMO) technologies, able to recover the modal cross-talk, allowed designers to reconsider the use of multimode fibers in order to add a further degree of data multiplexing in the space domain (SDM).

Single-mode fiber solitons and dispersion managed solitons had been seriously considered in the 1990s for developing high speed transport networks \cite{ESSIAMBRE2002232}. However, the advent of digital compensation convinced designers to use low-power transmission systems, where nonlinear effects were negligible, while linear detrimental effects were compensated electronically. 

Multimode fibers, however, are capable of transmitting several tens Watts of power, against less than 500 mW for a single-mode fiber; the propagation of solitons in MM fibers is therefore appealing for designing, for example, unrepeatered systems able to reach long distances (up to 400-500 km), thanks to the high peak power that multimode solitons can assume.

In this review work, we will retrace the steps that led to the discovery of solitons in multimode fiber, and will describe their peculiar properties such as modal condensation and pulse width invariance; we will see how those properties are essential for the design of SDM soliton systems. New details will be provided in the process of multimode soliton fission. A brief description will be given of the recently introduced thermodynamic theory of multimode soliton condensation.

\section{Linear and soliton propagation in multimode fibers}

A graded-index (GRIN) multimode fiber (MMF) is a cylindrical symmetric waveguide where the refractive index decreases with the radial distance $r$ as $n(r)=n_0 \sqrt{1-2\Delta(r/a)^\delta}$, being $n_0$ the index at the center of the core, $\Delta \approx (n_0-n_c)/n_0$ the relative index difference, $n_c$ the cladding index, $a$ the core radius and $\delta$ a shape factor \cite{agrawal_2023}.

In commercial parabolic OM4 GRIN fibers, it is $\delta \approx 2.08$; hence, the fiber assumes parabolic profile. This type of fiber is capable of propagating $Q$ groups of Laguerre-Gauss degenerate modes; each group includes $g_j$ modes and polarizations, $j=1,..,Q$; the number of modes and polarizations in the j-th group equals $g_j=2j$. The total number of propagated modes and polarizations is $M=Q(Q+1)=V^2/4$, with $V=n_0 k_0 a \sqrt{2\Delta}$ a normalized frequency, and $k_0=2\pi/\lambda$. The degenerate modes of a group have substantially equal propagation constant \cite{Olshansky:75} $\beta^{(j)} =n_0 k_0 \sqrt{1-2\Delta (j/Q)^{2\delta/(\delta+2)}}$, which are equi-spaced among modal groups. 

In commercial step-index MM fibers, the shape factor $\delta >> 10$, and the refractive index profile assumes rectangular shape; such fibers are capable of propagating $V^2/2$ modes whose field solution involves the Bessel functions \cite{agrawal_2023}; the propagation constants have non regular spacing in step-index fibers.

An optical field propagating in a MMF supporting $M$ modes can be expressed as 

\begin{equation}
	E(x,y,z,t) = \sum_{p=1}^{M}F_p(x,y)A_p(z,t)\exp{(in_0k_0z)},
\end{equation}

where $F_p(x,y)$ are the transverse field profiles of the modes and $A_p(z,t)$ the complex amplitudes, accounting for modal power, phase, modal dispersion and the temporal shape of the transmitted pulse.

In the case of parabolic GRIN fibers, the general wave equation for $E(x,y,z,t)$ can be solved under the paraxial and slowly varying envelope approximations, assuming $A_p(z,t)$ is a slowly varying function of $z$. This brings to the (3D+1) generalized NLSE model \cite{Hasegawa:80,agrawal_2023}, which propagates all modes on a single polarization component $A_x$ or $A_y$ of the total optical field; for polarization $x$ it is  (the dual is for $y$)

\begin{multline}
\frac{\partial A_x(x,y,z,t)}{\partial z} =\frac{i}{2k}\Big(\frac{\partial^2 A_x}{\partial x^2} +\frac{\partial^2 A_x}{\partial y^2}\Big) +i\sum_{n=2}^{4} \frac{\beta_n^{(x)}}{n!}\big(i\frac{\partial}{\partial t}\big)^n A_x-\frac{\alpha_x(\lambda)}{2}A_x+ \\
+\frac{ik}{2}\Big[\frac{n^2(x,y)}{n_0^2}-1 \Big]A_x+i\gamma\Big[(1-f_R)A_x\Big(\lvert A_x \rvert^2+\frac{2}{3}\lvert A_y \rvert^2 \Big)+ \\
+f_RA_p \int d\tau h_R(\tau) \Big(\lvert A_x(t-\tau) \rvert^2 +\frac{2}{3}\lvert A_y(t-\tau) \rvert^2 \Big) \Big] .
\label{eq:3Deq}
\end{multline}

In Eq. \ref{eq:3Deq}, $\beta_n^{(x)}$ is the n-th order chromatic dispersion term for polarization $x$, $\alpha_x(\lambda)$ the wavelength-dependent loss coefficient, $\gamma=n_2 \omega_0/(c \pi w^2)$ accounts for the Kerr nonlinearity, with $n_2$ (m$^2$/W) the nonlinear index coefficient and $w$ the beam waist, $h_R(\tau)$ is the Raman response \cite{Stolen1989}.

In the linear propagation regime, Eq. \ref{eq:3Deq} includes only the linear terms, and can be solved using the variational method \cite{Karlsson:92}, obtaining a propagating beam characterized by an effective waist $w(z)$ oscillating with distance $z$ around an input value $w_0$, according to the self-imaging law \cite{Hansson2020}

\begin{equation}
	w(z)=w_0 \Big[\cos{\Big( \frac{\pi z}{z_p} \Big)}^2+C\sin{\Big( \frac{\pi z}{z_p} \Big)}^2 \Big]^{1/2} ,
 \label{selfImaging}
\end{equation}

with $z_p=\pi a/\sqrt{2\Delta}$ the self-imaging period (typically of the order of 0.55 mm), and $C=z_p^2/(\pi \beta_0 a_0^2)^2$ the oscillation extent, being $a_0=w_0/\sqrt{2}$ a beam width; self-imaging is caused by the coherent interference among modes. 

\begin{figure}[h]
\includegraphics[width=0.8\textwidth]{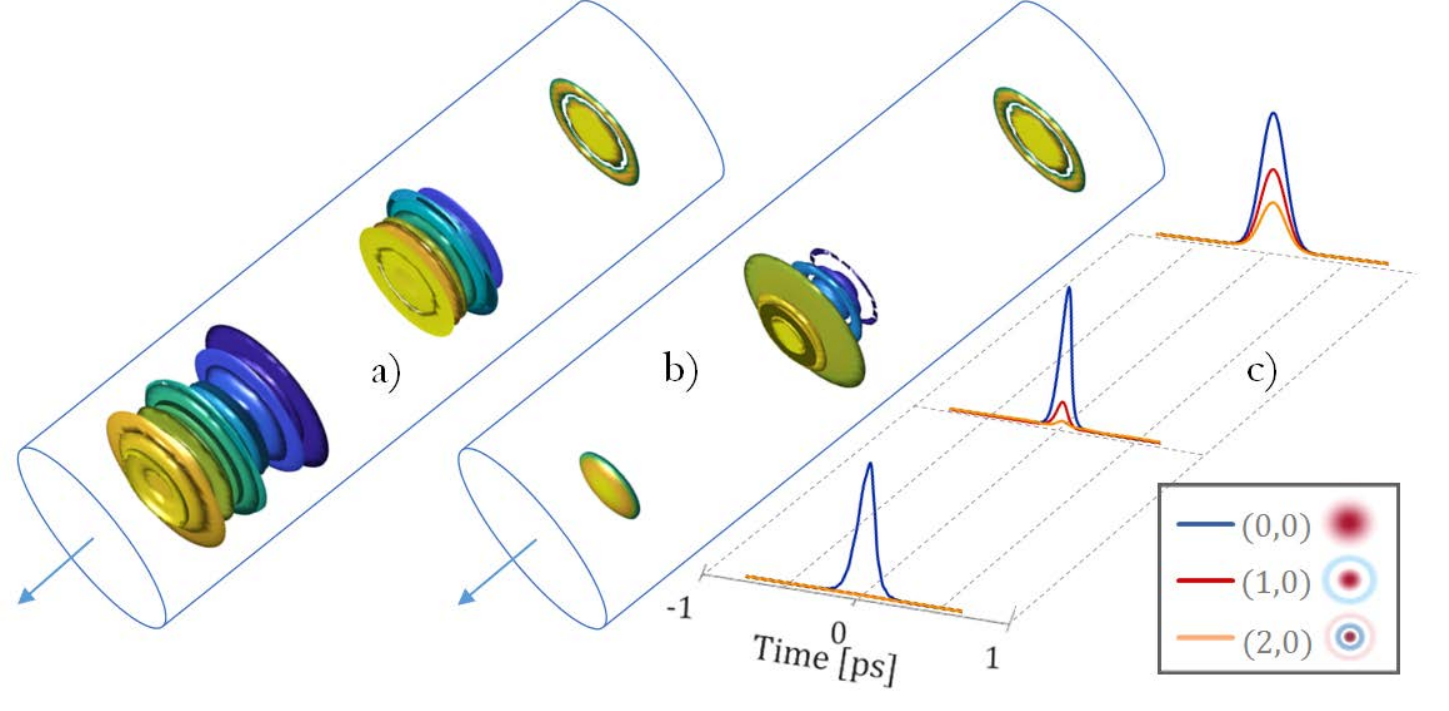}	\centering	
\caption{Numerical simulation, showing the propagation of a 3-mode bullet at the input and after 50 m and 1 km of GRIN fiber. a) 3D bullets in the linear regime, b) soliton regime, c) modal instantaneous power at soliton energy.}
\label{fig:Fig2_1}
\end{figure}

If a pulse is propagating instead of a continuous wave, the modal groups separate in time because of the modal dispersion. Modal separation is depicted in Fig. \ref{fig:Fig2_1}a, where 3 modes with axial symmetry, $LP_{01}$, $LP_{02}$ and $LP_{03}$ (sometimes referred to as (0,0), (1,0) and (2,0) when addressing to the radial and azimuthal indexes), are propagated over 1 km of GRIN fiber at low power; the 3 modes experience pulse broadening induced by the chromatic dispersion, as they separate.

The (3D+1) model of Eq. \ref{eq:3Deq} propagates all modes on a single polarization component $A_x$ or $A_y$ of the total optical field; 
it is very accurate in describing the propagation of a large number of modes, and it is also suitable for analyzing the propagation in fibers with complex index $n(x,y)$; however, it is not suitable for describing the process of multimode soliton formation, because it is lacking of the modal dispersion terms. 
As it will be shown in the following sections, multimode optical soliton propagation is characterized by an equilibrium among modal dispersion, Kerr and Raman nonlinearity and chromatic dispersion; for the correct analysis of soliton formation, it is of fundamental importance to include the effect of modal dispersion.

However, as a first approximation, we can consider the multimode soliton as already formed and apply a trial function to Eq. \ref{eq:3Deq} \cite{Renninger2013}

\begin{equation}
    E(x,y,z,t)=\sqrt{\frac{E_p}{2\pi w(z)^2 T_0}} \text{sech}\Big(\frac{t}{T_0}\Big) \exp{\Big[-\frac{x^2+y^2}{w(z)^2}+ i(\theta t^2+\alpha(x^2+y^2)+\phi)\Big]};
    \label{eq:3Dsolution}
\end{equation}

in Eq. \ref{eq:3Dsolution}, $w(z)$ is the beam waist, $E_p$ the pulse energy, $\theta, \alpha, \phi$ are chirp and phase parameters, and $T_0=T_{FWHM}/1.763$ is related to the full-width at half-maximum temporal pulse width. In \cite{RAGHAVAN2000377} and \cite{Ahsan:18} it was demonstrated that Eq. \ref{eq:3Dsolution} is a solution of the (3D+1) equation in the variational approximation, for negligible loss and Raman nonlinearity, and provided that the dispersion length $L_D=T_0^2/\lvert \beta_2 \rvert$, the nonlinearity length $L_{NL}=2T_0/(\gamma E_p)$, the diffraction length $L_{diff}=\beta_0 w_0^2$ and the self-imaging period $z_p$ are related as $L_D/L_{NL}=z_p/(\pi L_{diff}$) , with $\beta_0$ the propagation constant of the fundamental mode and $w_0$ the input beam waist, respectively. According to the solution of Eq. \ref{eq:3Dsolution}, the multimode soliton propagates with constant $sech$ temporal shape, and the beam waist undergoes periodic oscillations with distance, with period $z_p$, around the value $w_0$. 

\begin{figure}[h]
\includegraphics[width=0.5\textwidth]{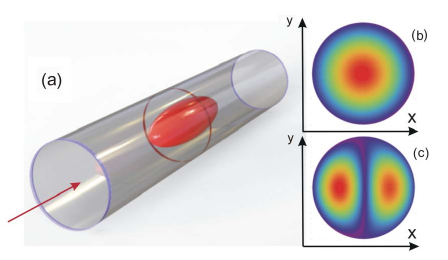}	
\centering	
\caption{Representation of stable spatiotemporal solitons propagating in a graded-index optical fiber. Two types of spatial cross-section profiles are shown on the right, namely, fundamental (b) and dipole-mode (c) solitons, respectively. With permission from ref \cite{PhysRevA.97.013841}, under a Creative Commons licence CC BY 4.0, American Physical Society.}
\label{fig:Fig2_2}
\end{figure}

The stability of spatiotemporal solitons in GRIN fibers was theoretically discussed and numerically analyzed in \cite{PhysRevA.97.013841}. It was observed that the effective two-dimensional potential formed by the graded refractive index prevents three-dimensional collapse into singularity as is known to occur in uniform three-dimensional media. Stable fundamental and dipole-mode spatiotemporal solitons were numerically found (Fig. \ref{fig:Fig2_2}).

A full description of the process of soliton formation must account for the delay among modal groups, or modal walk-off. A common numerical model accounting for this effect, used for pulse propagation in multimode fibers, is the coupled GNLSE \cite{Horak2012,Wright2018a} for mode and polarization $p$, modified to include modal and wavelength-dependent losses, and linear random-mode coupling (RMC) 

\begin{multline}
\frac{\partial A_p(z,t)}{\partial z} =i(\beta_0^{(p)}-\beta_0)A_p-(\beta_1^{(p)}-\beta_1) \frac{\partial A_p}{\partial t}+i\sum_{n=2}^{4} \frac{\beta_n^{(p)}}{n!}\big(i\frac{\partial}{\partial t}\big)^n A_p-\frac{\alpha_p(\lambda)}{2}A_p+ \\ 
+i\sum_m q_{mp}A_m+in_2k_0\sum_{l,m,n}S_{plmn} \big[(1-f_R)A_lA_mA_n^*+f_R A_l[h_R*(A_mA_n^*)]\big]  .
\label{eq:CoupledNLSE}
\end{multline}

In Eq.~\ref{eq:CoupledNLSE}, $\beta_n^{(p)}$ is the n-th order dispersion term (modal and chromatic) for mode $p$, $\alpha_p(\lambda)$ the modal and wavelength-dependent loss coefficient, $n_2$ (m$^2$/W) the nonlinear index coefficient multiplying the Kerr and Raman terms. $S_{plmn}$ is an overlap integral among modes, accounting for inter-modal four wave mixing (IM-FWM) and inter-mode stimulated Raman scattering (IM-SRS)

\begin{equation}
\begin{aligned}
		S_{plmn}= 	\frac{\iint F_p^*F_lF_mF_n^*\mathrm{d}x \mathrm{d}y}
	{\sqrt{\iint \lvert F_p \rvert^2 \mathrm{d}x \mathrm{d}y
            \iint \lvert F_l \rvert^2 \mathrm{d}x \mathrm{d}y
            \iint \lvert F_m \rvert^2 \mathrm{d}x \mathrm{d}y
            \iint \lvert F_n \rvert^2 \mathrm{d}x \mathrm{d}y}} .
	\end{aligned}	
\label{eq:NLcoeff}
\end{equation}

$q_{mp}$ is the linear RMC coupling coefficient, from mode $m$ to $p$, arising from randomly distributed index profile imperfections $\Delta n(x,y,z)$ \cite{HO2013491}

\begin{equation}
q_{mp}= 	\frac{k_0}{2 n_{m} (I_m I_p)^{1/2}}\iint \Delta n^2(x,y,z)F_p^*(x,y)F_m(x,y)\mathrm{d}x \mathrm{d}y ,
\label{eq:RMCcoeff}
\end{equation}

where $k_0$ is the vacuum wavenumber, $n_{m}$ the modal effective index, and $I_m, I_p$ the modal intensity transverse profile.
Degenerate modes are not accounted for in Eq.~\ref{eq:RMCcoeff}, because their coupling is so fast that power equipartition can be assumed into groups. 

Common parameters at $\lambda=1550$ nm for a GRIN OM4 MMF are: $\beta_2=-27$ ps$^2$/km, $\beta_3=0.14$ ps$^3$/km for the fundamental mode, $\alpha=0.21$ dB/km, negligible modal loss, $n_2=2.7\text{x}10^{-20}$ m$^2$/W,  $f_R=0.18$, index parabolic factor $g=2.08$, relative index difference $\Delta=0.010$, core radius $a=25 \mu$m.


\section{Modal attractors in multimode solitons}

The solution proposed in Eq. \ref{eq:3Dsolution} for a GRIN fiber, describing a self-imaging soliton which preserves the modal content over long distance, assumes that a multimode soliton has already formed, without considering the role of the modal dispersion. 

However, experimental results reported for MMF solitons reveal properties not accounted for by the periodical solution of Eq. \ref{eq:3Dsolution}.

A first experimental evidence of a multimode soliton generated by the stimulated Raman scattering (SRS) was reported in \cite{Grudinin:88}; an ultrashort pulse with 90 fs duration at 1650 nm was generated by injecting a 150 ps pump pulse at 1064 nm, with 600 kW peak power, into 500 m of parabolic GRIN fiber. 

In \cite{Renninger2013}, it was reported the generation of a multimode soliton at telecom wavelength. 300-fs pulses with energy up to 3 nJ were injected into 100 m of GRIN fiber; an output mode-field-diameter of 17.9 $\mu$m was obtained when injecting an 11.5 $\mu$m input beam; at 0.5 nJ input energy, it was obtained an output pulse width comparable to the input. For larger energies, it was observed a soliton affected by wavelength red-shift produced by the SRS; it was also reported a residual wave with no wavelength shift.

\begin{figure}[h]
\includegraphics[width=0.5\textwidth]{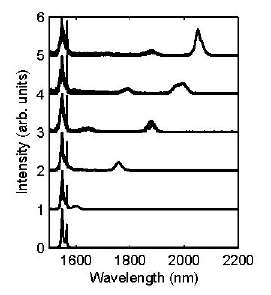}	\centering	
\caption{Experimental evidence of Raman soliton generation after 25 m of GRIN fiber. With permission from ref. \cite{Wright:15}, Optica Publishing Group.}
\label{fig:Fig4_1b}
\end{figure}

In \cite{Wright:15}, experimental results (Fig. \ref{fig:Fig4_1b}), compared to numerical simulations, confirmed the generation of Raman multimode solitons; 500-fs pulses at $\lambda=1550$ nm were launched into 25 m of GRIN fiber with $2a=62.5 \mu$m. The initial excitation was controlled in order to stimulate the higher-order modes at the input fiber end. Raman solitons were observed, affected by wavelength red-shift up to $\lambda=2100$ nm for input energy increasing from 3.2 nJ up to 34 nJ. The presence of a large amount of broadband dispersive wave \cite{Wright2015a}, related to the spatiotemporal oscillations of the soliton, was also detected.

In \cite{Wu:21}, numerical and experimental results reported the near-field, spectrum and pulse width of multimode solitons, when 500 fs pulses at 1550 nm were launched into a 15 cm GRIN fiber with a 50 $\mu$m core diameter and 0.2 numerical aperture. Spatial beam self-cleaning \cite{Krupa2017} was observed when there was significant overlap of the launched beam profile with the fundamental mode of the fiber. Output pulse width decreased from 480 fs to 30 fs when the input peak power was increased from 20 kW to 410 kW.

In \cite{Zhu:16}, it was reported the observation of multimode solitons in few-mode fibers (FMF), a GRIN fiber supporting only modes $LP_{01}$, $LP_{11a}$ and $LP_{11b}$. 280-fs pulses at $\lambda=1550$ nm were launched into 25 m of FMF at variable energy, controlling the input coupling so that the $LP_{01}$ mode (case IC1) or the $LP_{11}$ modes (case IC2, IC3) had larger power. Figures 4a and 4b in \cite{Zhu:16} show the simulated and measured output pulse width, respectively, for increasing input energy and in the three coupling conditions. Experimental data reported an output pulse width decreasing for growing input energy, converging to 120 fs for energy between 2 and 3 nJ. It will be shown in Sec. \ref{sec:Pulse width invariance} that this is a standard pulse width duration for multimode solitons at telecom wavelength.

In \cite{Zitelli:22} it was experimentally reported the generation of a soliton in a multimode step-index fiber; 70-fs pulses at 1450 nm were launched into 10 m of a commercial step-index fiber with $2a=50 \mu$m. A multimode soliton was observed at the output after injecting 21 nJ input energy. The soliton captured the modes $LP_{01}$, $LP_{11a}$ and $LP_{11b}$ while forming, and it evolved to a fundamental mode Raman soliton; the residual wave included the higher-order modes, and was lacking of the fundamental mode.


\FloatBarrier


In \cite{Zitelli2021} and \cite{Zitelli2021a} they were reported numerical simulations and experimental results using GRIN fibers, describing a complex scenario for soliton formation, where the modal walk-off length $L_W=T_0/\Delta \beta_1$ comes into play; the quantity

\begin{equation}
    \Delta \beta_1=\frac{\sum_j P_j(\beta_{1j}-\beta_{10})}{\sum_j P_j} ,
    \label{eq:meanBeta1}
\end{equation}

in (s/m), is the differential first-order dispersion coefficient with respect to the fundamental mode, averaged among the propagating modes with power $P_j$.

\begin{figure}[h]
\includegraphics[width=0.8\textwidth]{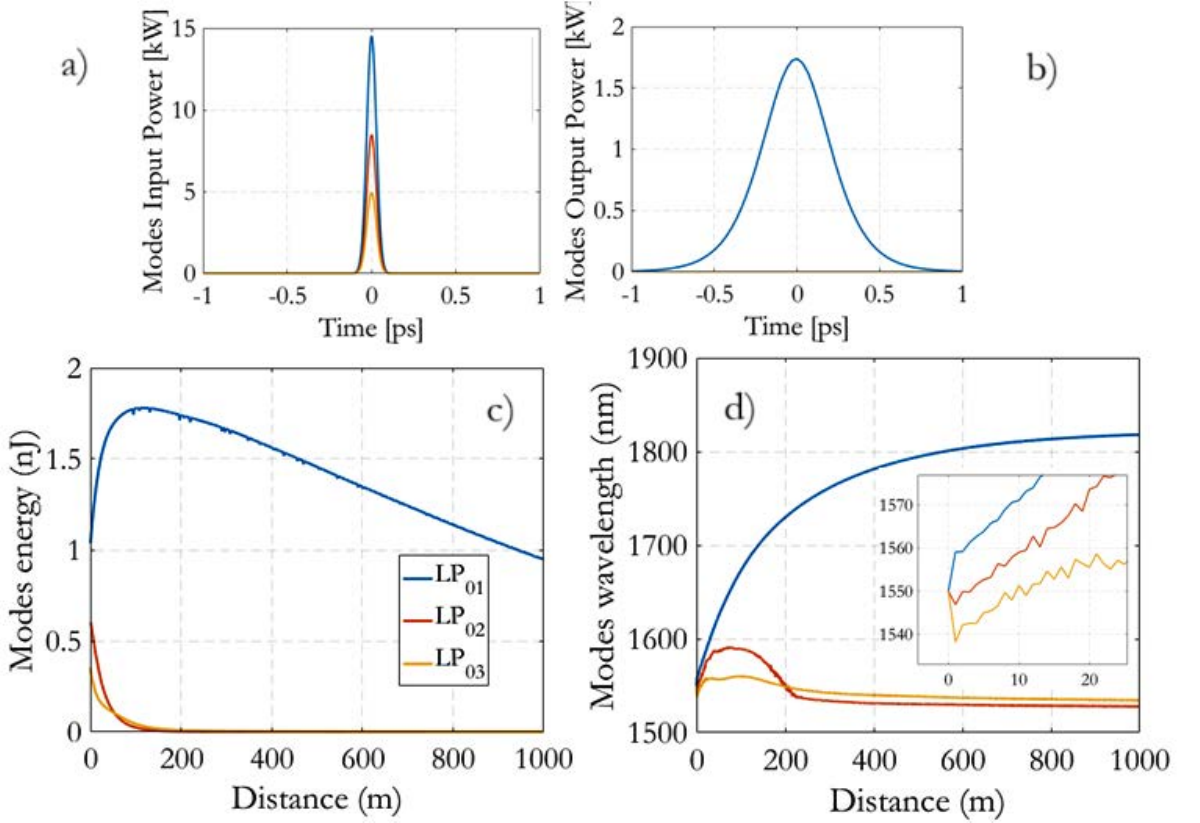}	\centering	
\caption{Numerical simulation, showing a 70-fs pulse at $\lambda=1550$ nm and 28 kW peak power, including 3 radial modes propagating over 1 km of GRIN OM4 fiber. a), b): modal instantaneous power at the input and output, respectively. c), d): Modal energy and wavelength vs. distance, respectively. With permission from ref. \cite{Zitelli2021}, Optica Publishing Group.}
\label{fig:Fig4_2}
\end{figure}

Figure \ref{fig:Fig4_2} shows a numerical simulation performed in \cite{Zitelli2021} using the model of Eq. \ref{eq:CoupledNLSE}, where a 70-fs pulse at $\lambda=1550$ nm is launched at the input of a GRIN OM4 fiber with 3 radial modes $LP_{01}$, $LP_{02}$ and $LP_{03}$, at a total peak power $P_0=28$ kW. Modal pulses form a multimode soliton after several meters distance. The power profile of the 3 modes at the fiber input end is represented in Fig. \ref{fig:Fig4_2}a. During the first few meters of propagation, the Kerr nonlinearity causes a wavelength shift of the spectra of the modes (inset of Fig. \ref{fig:Fig4_2}d), in such a way that the chromatic dispersion compensates for the modal dispersion. Once the modal velocities are equalized, the 3 overlapping pulses propagate exchanging power by IM-FWM and IM-SRS; the nonlinear process promotes the fundamental mode, which acts as an attractor the energy of the other modes \cite{YANKOV1997147} (Fig. \ref{fig:Fig4_2}c). 

Figure \ref{fig:Fig2_1}b depicts the propagating bullets in the soliton regime, as power is gradually transferred to the fundamental mode. Figure \ref{fig:Fig2_1}c shows the temporal profiles of the 3 propagating modes. The basic difference respect to the linear regime of Fig. \ref{fig:Fig2_1}a, is that the mode groups do not separate in time, and propagate while mutually trapped; once the mode $LP_{01}$ has attracted all the pulse energy, the soliton propagates as a fundamental mode bullet.

At the same time, stimulated Raman scattering induces a wavelength red-shift to the modal pulses (Fig. \ref{fig:Fig4_2}d), or soliton self-frequency shift (SSFS), whose change per unit distance can be quantified as \cite{Gordon1986, Zitelli_9887813}

\begin{equation}
    \frac{d\Delta\lambda}{dz}=\frac{4T_R}{15\pi c\lambda^2(z) \lvert \beta_2(z) \rvert ^3} \Big( \frac{n_2 E_{s0}}{w_e^2} \Big)^4 ,
    \label{eq:RamanShift}
\end{equation}

with $w_e$ the effective beam waist, and $T_R=f_R\int\tau h_R(\tau)d\tau$ the mean Raman response time, equal to approximately 3 fs in silica fibers. Experimental evidence of the SSFS is given in Fig. \ref{fig:Fig4_1b}.

Since the higher-order modes have lower wavelength with respect to $LP_{01}$, the IM-SRS boosts the transfer of power to the fundamental mode. After 100 m distance, nearly all the energy has been transferred to the $LP_{01}$ mode, which propagates as a fundamental mode Raman soliton \cite{Zitelli2021}. Fiber loss reduces the pulse energy; as a consequence, the soliton must increase the pulse width according to the soliton equation relating the pulse energy and the pulse width \cite{Renninger2013}

\begin{equation}
    E_p=\frac{\lambda \lvert \beta_2(\lambda) \rvert w_e^2}{n_2 T_0} .
    \label{eq:solitonEnergy}
\end{equation}

After 1 km propagation, a single-mode larger soliton remains (Fig. \ref{fig:Fig4_2}b) with some residual wave still conserving the input wavelength.

\FloatBarrier

\begin{figure}[h]
\includegraphics[width=0.95\textwidth]{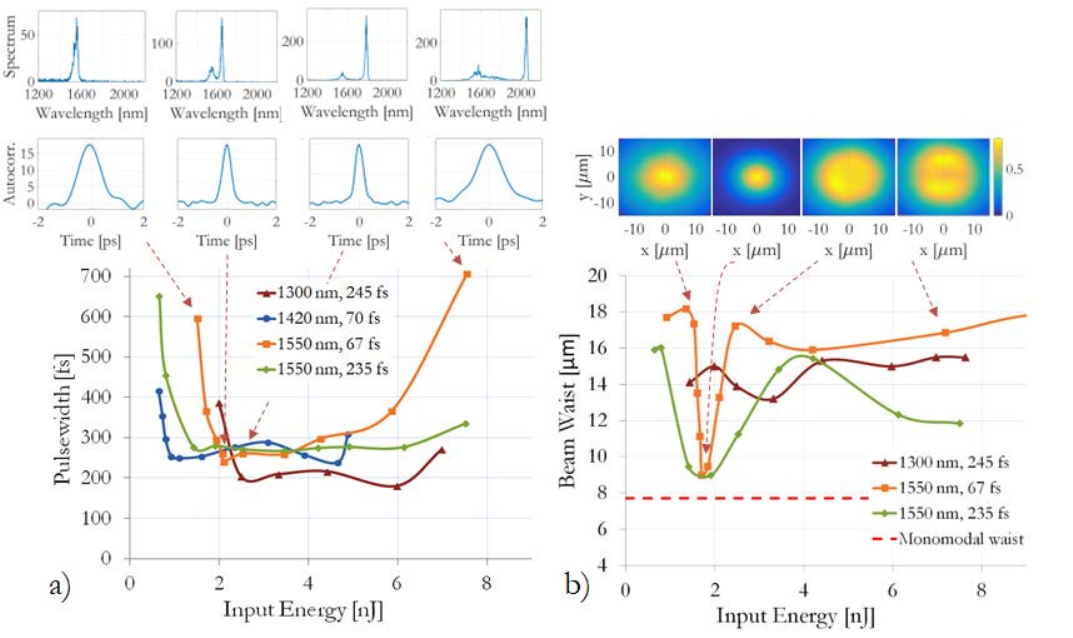}	\centering	
\caption{Experimental results propagating 67 or 245 fs pulses at $\lambda=1300$ to $1550$ nm over 120 m of GRIN OM4 fiber. a), b): output pulse width and beam waist at 120 m distance, respectively. The insets above the plots are the measured spectra, autocorrelation traces and beam near-filed, at $\lambda=1550$ and at the indicated pulse energies. Adapted with permission from ref. \cite{Zitelli2021a}, under a
Creative Commons licence CC BY 4.0, Springer Nature Limited.}
\label{fig:Fig4_3}
\end{figure}

Figure \ref{fig:Fig4_3} shows the experimental results from \cite{Zitelli2021a}, when propagating 67 or 245 fs pulses at $\lambda=1300$ to $1550$ nm over 120 m of GRIN OM4 fiber. The insets on the top of Fig. \ref{fig:Fig4_3}a illustrate the measured autocorrelation traces and spectra for input pulse width 67 fs and wavelength 1550 nm. For increasing input pulse energy $E_p=1$ to $8$ nJ, the output pulse width has a minimum of 260 fs at an optimal energy of 2 nJ. Correspondingly, the soliton pulse wavelength experiences an increasing red-shift with respect to the small residual wave, whose wavelength is conserved. Figure \ref{fig:Fig4_3}a also shows that the minimum pulse width does not change significantly for different input pulse width or wavelength, with the exception of $\lambda=1300$ nm, near the zero-chromatic dispersion wavelength. 

The insets above Fig. \ref{fig:Fig4_3}b illustrate the corresponding near-fields measured for a 67 fs pulse at 1550 nm. For increasing energy, the output waist reduces as the soliton is forming; at an optimal 2 nJ input energy, the output beam waist approaches to the theoretical value of the fundamental mode ($w_0=7.8 \mu$m); for higher energies, the output beam waist assumes larger values, denoting the presence of a large amount of dispersive wave spatially overlapped to the soliton near-field. When a different input pulse width (245 fs) is tested at the same wavelength, the same values of the case with 67 fs input pulse width are found for the minimum waist and the corresponding energy. The 2 nJ energy value corresponding to minimum waist was also found for the minimum pulse width.

The experiment confirms the dynamics of soliton formation already seen in Fig. \ref{fig:Fig4_2}. As a consequence of IM-FWM and IM-SRS, multimode soliton evolves, at an optimal energy, to a fundamental mode soliton experiencing Raman red-shift; a residual wave is eventually observed, which is generally minimized at an optimal input energy.

\FloatBarrier

\section{Pulse width invariance} \label{sec:Pulse width invariance}

\begin{figure}[h]
\includegraphics[width=0.7\textwidth]{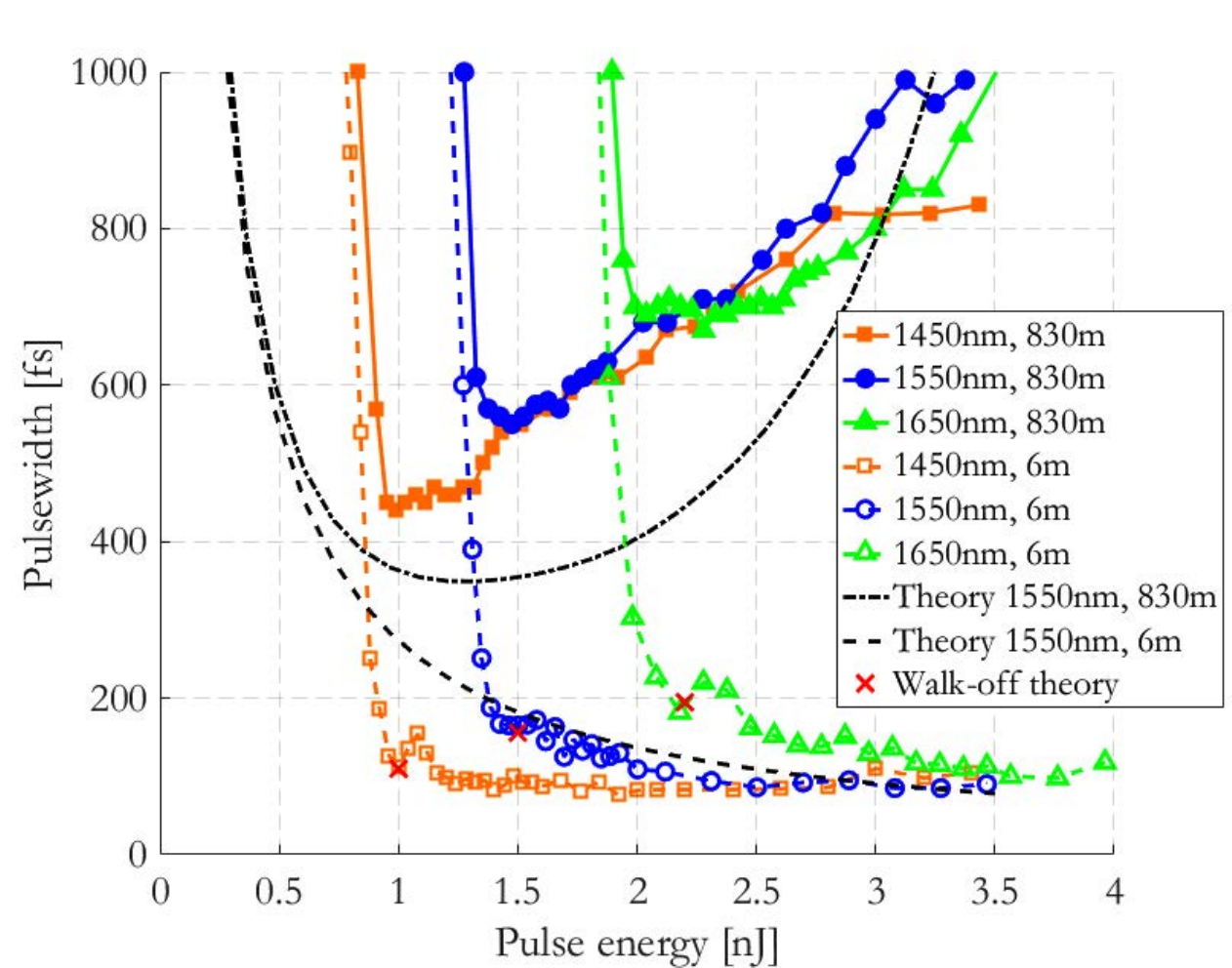}	\centering	
\caption{Experimental pulse width vs. soliton energy for a pulse propagating over 6 m or 830 m of GRIN fiber. Input parameters are: 1450 nm and 60 fs, 1550 nm and 70 fs, 1650 nm and 60 fs; 30 $\mu$m beam diameter. Red marks are calculated from Eq. \ref{eq:solitonPulsewidth}. With permission from ref. \cite{Zitelli_9887813}, Optica Publishing Group.}
\label{fig:Fig5_1}
\end{figure}

In  \cite{Zitelli_9887813}, the experimental data of Fig. \ref{fig:Fig5_1} were reported, which must be compared with the experimental results of Fig. \ref{fig:Fig4_3} in order to have a better insight into the process of multimode soliton formation.

In the figure, it is shown the output pulse width vs. pulse energy, measured at $\lambda=1450, 1550$ or $1650$ nm for a 70 fs input pulse width, over different lengths of GRIN OM4 fiber. At 6 m distance, pulse width rapidly reaches a minimum of 100 fs for increasing energy; it is not possible to find, at this short distance, an optimal soliton energy, because it has not fully formed yet.

For increasing length, the existence of an optimal soliton energy becomes clear; at 830 m distance, a sharp minimum of the output pulse width is found at $E_s=1.5, 2$ or $2.5$ nJ input energy for the 3 wavelengths, respectively. Hence, the initial pulsewith $T_{s0}$ of the soliton must be measured at short distance, in correspondence to the optimal energy measured after long distance. In the specific case of Fig. \ref{fig:Fig5_1}, it is $T_{s0}=70, 120$ and $170$ fs for the three wavelengths.

\FloatBarrier

\begin{figure}[h]
\includegraphics[width=0.7\textwidth]{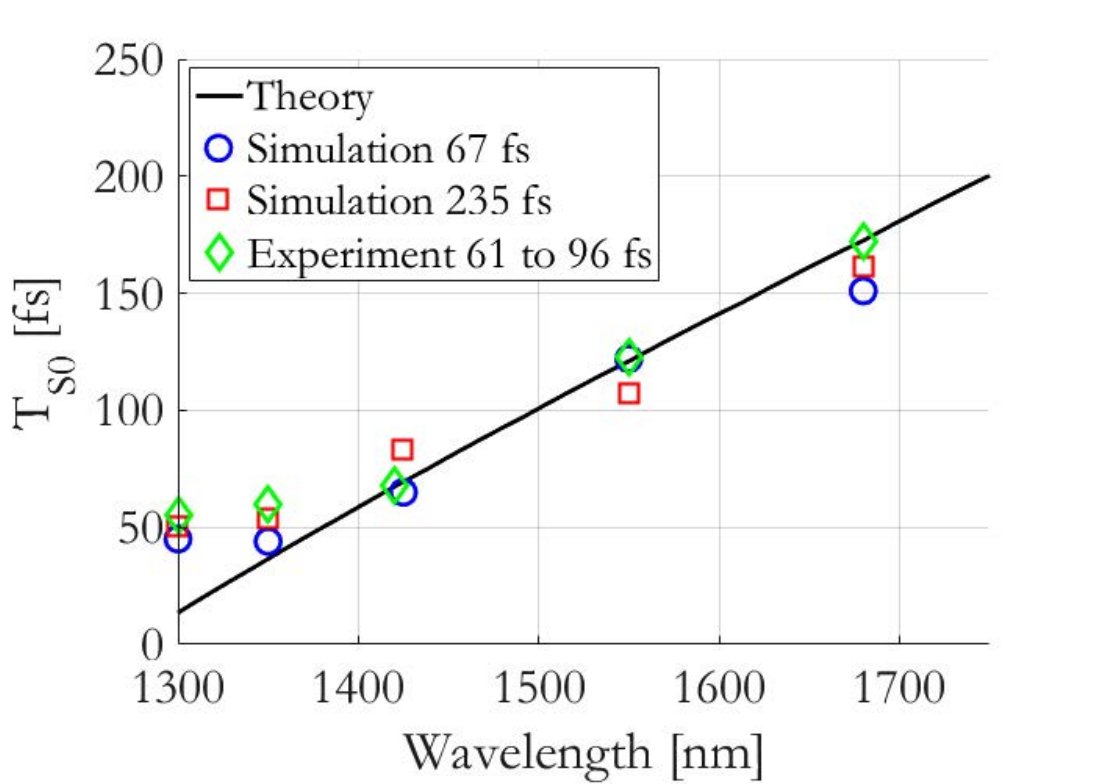}	\centering	
\caption{Theoretical curve of soliton pulse duration vs. wavelength, compared with measured and simulated soliton pulse width. Experiments carried out with 1 m of GRIN fiber, with input wavelength and pulse width: 1300 nm and 61 fs, 1350 nm and 61 fs, or 1420 nm and 70 fs, respectively. Experiment carried out with 6 m of GRIN fiber, with input wavelength and pulse width: 1550 nm and 67 fs, 1680 nm and 96 fs, respectively. With permission from ref. \cite{Zitelli2021a}, under a
Creative Commons licence CC BY 4.0, Springer Nature Limited.}
\label{fig:Fig5_2}
\end{figure}

In \cite{Zitelli2021a}, soliton propagation was experimentally investigated with cut-backs from 120 m to 1 m of GRIN fiber, for different input pulse width and wavelength; the experimental data of Fig. \ref{fig:Fig5_2} were obtained. Here, the initial pulse width of the formed soliton $T_{s0}$ is reported vs. the input wavelength for different input pulse widths, which do not seem affecting the value of $T_{s0}$. Experimental data were collected at the optimal soliton energy $E_{s0}$ corresponding to minimum long-distance pulse width; we recall that this is generally larger that $T_{s0}$, because of linear losses and the red-shift induced by the Raman soliton self-frequency shift (SSFS) \cite{Grudinin:88, Gordon1986}. 

In the same figure, numerical simulations obtained using the model of Eq. \ref{eq:CoupledNLSE} are reported.
Simulations show that two pulses with different input pulse width (67 or 235 fs), and same wavelength (1350, 1550 or 1680 nm), produce the same multimode soliton after 3 to 6 m distance. The pulse width $T_{s0}$ immediately after the soliton formation looks affected only by the input wavelength. 


This property of pulse width invariance appears as a discriminating point for multimode solitons with respect to the single-mode counterpart \cite{Zakharov1970ExactTO, Agrawal2013}. Here is where the modal walk-off length $L_w$ comes into play. Solitons in single-mode fiber form when the chromatic dispersion length $L_D$, measuring the distance where pulses have broadened by a factor $\sqrt{2}$, balance the nonlinearity length $L_{NL}$, indicating where the nonlinear phase shift induced by the self-phase modulation (SPM) equals 1 rad \cite{Agrawal2013}. 

In multimode fibers, other distances to consider are the random mode coupling and birefringence correlation lengths, $L_{cm}$ and $L_{cp}$, defined as the characteristic lengths associated with the linear coupling between modes or between polarizations, respectively \cite{HO2013491, Xiao2014TheoryOI}.

In order to explain the observations of Fig. \ref{fig:Fig5_2},
it must be assumed that nonlinearity acts over distances shorter than those associated with random mode coupling and birefringence, i.e., for $L_{NL} < L_{cm}, L_{cp}$. Besides, in multimode solitons the modal dispersion must balance the chromatic dispersion as well; modal pulses need to propagate at the same speed before interacting by IM-FWM and IM-SRS; hence, the modal walk-off length must be comparable to the dispersion and nonlinearity lengths.
 A condition for temporal trapping of the optical modes was originally predicted, for the j-th mode, as \cite{Hasegawa:80, Crosignani:81} $L_{wj} \geq L_D=L_{NL}$. This condition was restricted in \cite{Zitelli2021a} to $L_D=L_{NL}=\delta L_w$, with $\delta$ an experimental adjustment constant close to unity. By replacing the respective expressions, it was found for the initial soliton pulse width

\begin{equation}
    T_{s0}(\lambda)=\delta \cdot 1.763 \frac{\lvert \beta_2(\lambda) \rvert}{\Delta \beta_1(\lambda)} .
    \label{eq:solitonPulsewidth}
\end{equation}

Equation \ref{eq:solitonPulsewidth} fairly predicts the value of multimode solitons pulse width, given the wavelength and the fiber dispersion parameters. Because of the fundamental role of the walk-off length in the process, the name of walk-off solitons was proposed for this type of pulses. In Fig. \ref{fig:Fig5_2}, the black curve is calculated from Eq. \ref{eq:solitonPulsewidth}, showing good agreement with the experimental and numerical data. In Fig. \ref{fig:Fig5_1} the theoretical soliton pulse width from Eq. \ref{eq:solitonPulsewidth} are reported as red marks over the short distance curves, for the three tested wavelengths.

According to the observations of Figs. \ref{fig:Fig5_1} and \ref{fig:Fig5_2}, 
the walk-off solitons form at a specific pulse energy calculated from Eqs. \ref{eq:solitonPulsewidth} and \ref{eq:solitonEnergy} (with $T_0=T_{s0}/1.763$). For a GRIN OM4 fiber at $\lambda=1550$ nm, it is $T_{s0}=120$ fs and $E_s=1.5$ nJ, plus an amount of input energy eventually remaining in the residual wave. 

The invariance property for the pulse width and the energy of a multimode soliton implies that, when injecting a pulse with proper energy $E_{s0}$ and pulse width larger or shorter than $T_{s0}$, the travelling modes adjust their pulse width to the soliton value, Eq. \ref{eq:solitonPulsewidth}, depending on the fiber dispersion and the input wavelength. In the case of input pulse width much larger than $T_{s0}$ (for example, few picoseconds at $\lambda=1550$ nm), the multimode soliton does not form; this property is important for the design of soliton space-division multiplexed systems (SDM), as it will be seen in Sec. \ref{sec:Optical solitons in SDM systems}. If, on the other hand, the input energy is much larger than $E_{s0}$, a soliton pulse may originate with proper pulse width and energy, emerging from a large amount of dispersive wave.

\section{Soliton fission and interaction}

In those cases where the modal content of the input pulse is complex, for example if the higher-order modes possess larger energy respect to the fundamental, or a large number of modes is propagating, or the input pulse energy is greater than $E_{s0}$ from Eq. \ref{eq:solitonEnergy}, multiple solitons can originate from a single input pulse.

In \cite{Wright:15}, experimental evidence of multiple soliton generation was provided using 95 m of GRIN fiber, and launching 500-fs pulses at $\lambda=1550$ nm. A main higher energy Raman soliton was observed at the output, with red-shifted 1620 nm wavelength and 10 $\mu$m beam waist. A smaller secondary soliton spectrum was also detected, together with a large amount of dispersive wave.

Cherenkov spectral peaks, generated by  multimode soliton fission processes in nonlinear MMFs, were observed in \cite{Eftekhar2021}. It was reported the formation of a multimode soliton, that forces all the modes involved to first coalesce in the temporal domain, and in doing so, individually shift their spectral content. A phase-matching condition of solitons after fission occurs, which is the cause of the Cherenkov dispersive waves extending from the 1550 nm input wavelength down to the normal dispersion region of a MMF.

In \cite{Buch:16,Sun:22} it was numerically and experimentally investigated the collision interaction of multimode solitons. When propagating solitons into degenerate modes, it was found that each soliton transfers some of its power to the other mode through IM-FWM. In spite of this intermodal power transfer, each soliton keeps propagating as a bimodal soliton and interacts with the other bimodal soliton as if they were propagating inside a single-mode fiber. The collision outcome is dependent on the input relative phase of the two modes. Fig. 3 of \cite{Buch:16} shows the collision between modes $LP_{11a}$ and $LP_{11b}$ with different values of relative phase between the two input beams; a complete transfer of power between the two modes is observed for relative phase $\pi/8$, which reduces to a partial transfer for phase $\pi/4$ and is absent for phase $\pi/2$.


In \cite{Zitelli:20}, it was numerically and experimentally investigated the soliton fission in short spans (30 cm) of GRIN fiber at the telecom wavelength. For input energy larger than 15 nJ, the presence of nonlinear losses had to be taken into account. In this regime of extreme nonlinear effects, it was observed the generation of a train of solitary pulses with common pulse width (50-60 fs); pulse duration was observed to be nearly constant for input energy in the range 15-50 nJ.

\begin{figure}[h]
\includegraphics[width=0.9\textwidth]{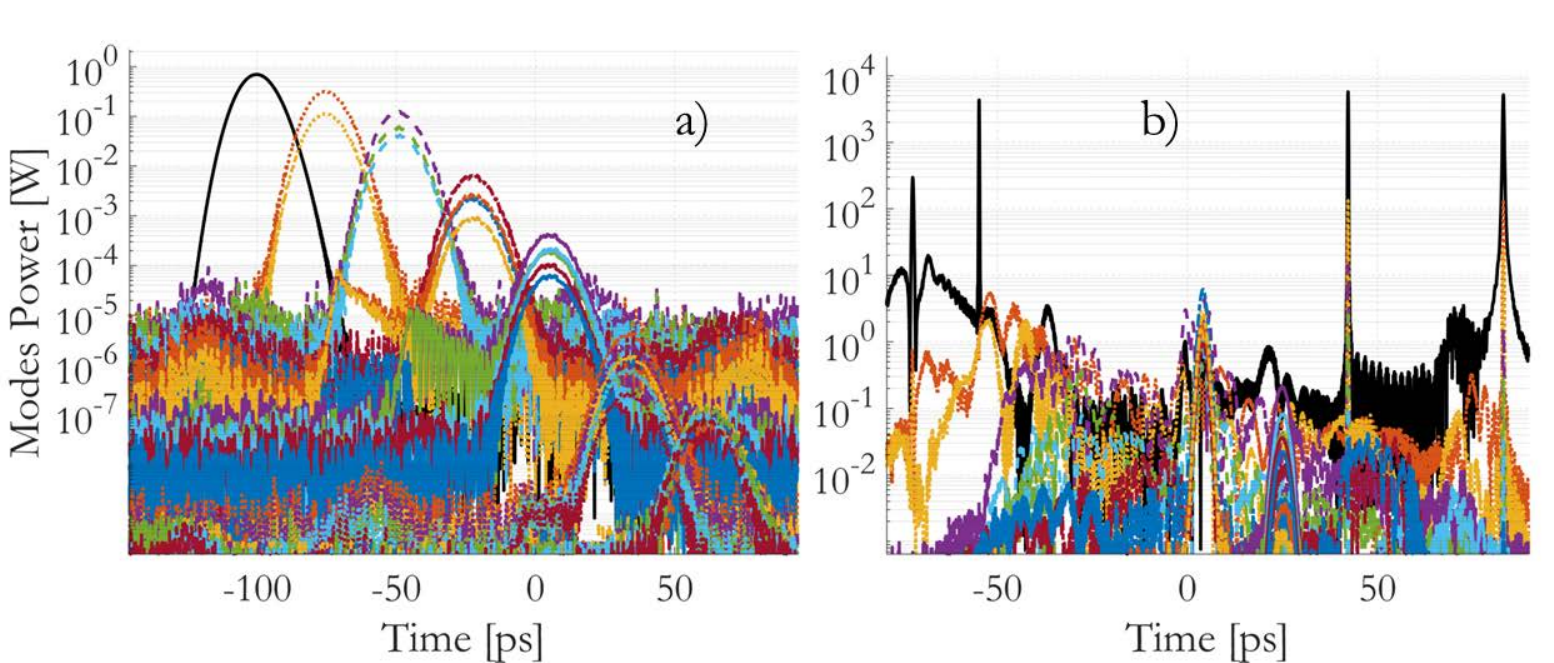}	\centering	
\caption{Modal instantaneous power for a 250 fs pulse at 1400 nm, propagating in a long span of GRIN fiber; pulse is composed by 28 Laguerre-Gauss modes with uniform power distribution at the input. a): linear regime, b): soliton power.}
\label{fig:Fig6_2}
\end{figure}

Recently, it was proposed an alternative interpretation of the fission mechanism for multimode solitons, involving the interplay of modal dispersion, random-mode coupling (RMC) and IM-FWM \cite{zitelli2023statistics}. The process is described by Fig. \ref{fig:Fig6_2} showing a numerical simulation, performed using the model of Eq. \ref{eq:CoupledNLSE}, of a 250 fs pulse at 1400 nm, propagating in a long span of GRIN fiber; pulse is composed by 28 Laguerre-Gauss modes having uniform power distribution at the input. 
Figure \ref{fig:Fig6_2}a illustrates the propagation with negligible nonlinearity (input energy 0.02 nJ); pulses composed by the individual groups of degenerate modes separate in time and broaden as a consequence of modal and chromatic dispersion, respectively. If RMC was not present, the modal content of the pulses would be limited to the respective groups; instead, RMC transfers energy to the other groups, in a way that each pulse will also include modes from the lower-order and higher-order groups and from the fundamental mode (the black curve in the figure); the result is a multimodal background noise into each propagating pulse. 
When the input pulse is increased to soliton level (5.0 nJ in Fig. \ref{fig:Fig6_2}b), the self-phase modulation (SPM) counteracts the dispersion-induced broadening; IM-FWM transfers power among the modes of a same pulse; the fundamental mode, which is present into each pulse as a consequence of the RMC, is an attractor from the higher-order modes; after hundreds meters distance, a train of fundamental mode solitons is obtained, each one assuming the pulse width given by Eq. \ref{eq:solitonPulsewidth}. Solitons are affected by Raman self-frequency shift, and IM-SRS provides a contribution to the process of condensation to the fundamental mode.

\FloatBarrier

\section{Thermodynamic approach of multimode solitons}

Optical thermodynamics of multimode systems \cite{Wu2019,doi:10.1080/23746149.2023.2228018,pourbeyram2022direct} is a powerful tool for analyzing complex multimode interactions. When a physical process is responsible for a chaotic power transfer among fiber modes, such as the IM-FWM, and provided the pulse energy and peak power are preserved, the power distribution at the output of a multimode fiber can be described by a Rayleigh-Jeans or by a Bose-Einstein law.

Multimode soliton propagation is an example of energy and peak power conservation. It was experimentally demonstrated \cite{zitelli2023statistics} that the process of multimode soliton generation can be described as a modal condensation in thermodynamic terms and, during its formation, the modal power content assumes a Bose-Einstein distribution. 
250-fs pulses at 1400 nm were launched  into 830 m of OM4 GRIN fiber; the power distribution of the modal groups was measured starting from the peaks of the detected pulses, corresponding to the different modal groups, similar to what is shown in Fig. \ref{fig:Fig6_2}. It was found that, at soliton power, the mean modal power fraction of the degenerate modes with same eigenvalue (the differential propagation constant $\epsilon_i=\beta_i-\beta_M, i=1,..,M$ , being $M$ the number of propagating modes) followed a Bose-Einstein distribution when plotted against $\epsilon_i$.
More than half of the total power was attracted by the fundamental mode (the first detected pulse). The corresponding output near field showed the fundamental mode dominating over a residual of higher-order modes. Consequently, it was found that the output mean modal power fraction condensates to the fundamental mode. 
The Bose-Einstein law fitted the modal distribution up to the 9-th modal group (45 modes per polarization), demonstrating that the thermodynamic approach is an effective way to describe the process of multimode soliton formation and condensation to the fundamental mode.

\FloatBarrier

\section{Optical solitons in SDM systems} \label{sec:Optical solitons in SDM systems}

The inability of picosecond pulses to form a multimode soliton, discussed in Sec. \ref{sec:Pulse width invariance}, is the basis of soliton SDM, where the individual channels are coupled into single modes or groups of degenerate modes. In \cite{Mecozzi:12} it was demonstrated that, although using large pulses, individual groups of degenerate or quasi-degenerate modes are able to generate intra-group Manakov solitons; RMC among degenerate modes, which is responsible for a disordered transfer of energy among the modes of a same group, does not prevent the formation of those solitary solutions predicted by the coupled Manakov equations \cite{Mumtaz:13}; to the contrary, disorder causes a fast energy exchange among degenerate modes, which minimize the modal dispersion within one group and promotes the formation of Manakov solitons.

By properly distributing the input power between the mode groups, one soliton channel for each group can be generated \cite{10034667}; the solitons of different groups cannot merge into a single multimode soliton, because the input pulse width does not satisfy the walk-off condition represented by Eq. \ref{eq:solitonPulsewidth}. Hence, soliton channels propagate at different velocities, and experience elastic collisions with the other modal groups in a similar way to what it is observed in wavelength-division multiplexed systems. This property permits the design of stable, high-speed multimode soliton SDM systems.

\section{Conclusions}

Multimode fiber solitons demonstrate unique properties, such as pulse width invariance and energy condensation, which are not found in the single-mode counterpart. When picosecond pulses are propagated, compatible with telecom applications, the disorder caused by fiber imperfections promotes the formation of Manakov solitons within the individual modal groups. However, the picosecond soliton pulses generated within the groups are inhibited from merging into a single multimode soliton, because Eq. \ref{eq:solitonPulsewidth} is not satisfied. Hence, stable modal soliton channels are possible, performing elastic collisions with other channels; this lays the foundation for the design of soliton SDM transmission systems.

The ability of MM fibers to transport tens of Watt of optical power, makes it possible to design high peak power soliton systems, between a few Watts and a few tens of kW, able to propagate over long distance with no amplification, or using Raman amplification pumped from the transmitter or the receiver side. Unrepeatered systems of this type can potentially double the distance that is commonly obtained with single-mode fiber.

Furthermore, the capability of condensating all the launched power into the fundamental mode of a MM soliton, improves the quality of beam power delivery systems employing multimode fibers.

\section{Backmatter}

\begin{backmatter}

\textbf{Funding} This work was supported by: 

Project ECS 0000024 Rome Technopole, Funded by the European Union - NextGenerationEU.


\noindent \textbf{Disclosures} The author declares no conflicts of interest.


\smallskip

\end{backmatter}








\end{document}